\documentclass[preprint,nofootinbib,aps,superscriptaddress,eqsecnum]{revtex4-1} 
 \pdfoutput=1
\usepackage{epsfig} 
\usepackage{amsmath} 
\usepackage{physics}
\usepackage{amssymb}
\usepackage{graphicx}
\usepackage{color}
\usepackage{float}
\usepackage{soul,xcolor}
\usepackage[font=scriptsize]{caption}
\usepackage{braket}
\usepackage{bbold}
\usepackage{subfigure}
\usepackage{braket}
\usepackage{titlesec}
\usepackage{placeins}
\usepackage{hyperref}
\usepackage{caption}
\usepackage{array}
\usepackage{booktabs}
\usepackage{subcaption}
\usepackage{changepage}
\usepackage{gensymb}
\usepackage{trimclip}
\usepackage{accents}
\usepackage{booktabs}
\usepackage{multirow}
\usepackage{lipsum}
\usepackage{placeins}
\usepackage{comment}

\PassOptionsToPackage{unicode}{hyperref}
\PassOptionsToPackage{naturalnames}{hyperref}

\usepackage[justification=centering]{caption}
\newcolumntype{P}[1]{>{\centering\arraybackslash}p{#1}}
\hypersetup{colorlinks,linkcolor={blue},citecolor={blue},urlcolor={blue}}

\def\bea{\begin{eqnarray}}
\def\eea{\end{eqnarray}}
\def\be{\begin{equation}}
\def\ee{\end{equation}}

\begin{document}

\title{Study of quantum decoherence at Protvino to ORCA experiment}

\author{Chinmay Bera}
\email[Email Address: ]{chinmay20pphy014@mahindrauniversity.edu.in}
\affiliation{Department of Physics, \'Ecole Centrale School of Engineering - Mahindra University, Hyderabad, Telangana, 500043, India}

\author{K. N. Deepthi}
\email[Email Address: ]{nagadeepthi.kuchibhatla@mahindrauniversity.edu.in}
\affiliation{Department of Physics, \'Ecole Centrale School of Engineering - Mahindra University, Hyderabad, Telangana, 500043, India}

\begin{abstract}
Protvino to ORCA (Oscillation Research with Cosmics in the Abyss) (P2O) is an upcoming neutrino oscillation experiment with a very long baseline of 2595 km. Due to the substantial baseline, this experiment provides a unique opportunity to study the earth matter effects over very large distances. This makes it a suitable experiment to investigate the environmental decoherence in neutrino oscillations, where the neutrino system could interact with a stochastic environment and lead to a loss in the coherence of neutrino states. 
In this work, we consider an open quantum system framework to simulate the neutrino oscillations in P2O experiment and obtain bounds on the decoherence parameters in different phenomenological models. Here we also consider that the decoherence parameter $\Gamma$ depends on neutrino energy $E_\nu$ as $\Gamma_{jk}(E_\nu) = \Gamma_{0} (\frac{E_\nu}{E_0})^n$. 
Further, assuming the presence of decoherence in nature we study the effect of it on the determination of neutrino mass hierarchy and CP violation at P2O experiment.
\end{abstract}


\maketitle

\section{Introduction}
Over the past few years, leading neutrino experiments have been diligently collecting data to estimate the neutrino oscillation parameters precisely. Till date, the majority of the data from these experiments aligns with the conventional three flavour neutrino oscillation framework, characterized by the three mixing angles, two mass square differences, and a Dirac Charge-Parity (CP) phase. Nevertheless, the forthcoming high precision neutrino oscillation experiments provide a significant opportunity to probe sub-leading new physics (NP) phenomenon. One such interesting phenomenon is the environmentally induced decoherence among the neutrino states passing through the earth matter. 

The standard three flavour neutrino oscillation phenomenology considers neutrinos to be isolated from the environment and the coherence of the neutrino states is preserved over very long distances. However several theories ~\cite{Hawking:1978sf, Amati:1987string, Saurya:2018se, Stuttard:2020mik, Luciano:2021qg} propose the possibility of interaction of neutrinos with a stochastic environment, leading to the loss of coherence in the neutrino states. This phenomenon is called quantum neutrino decoherence~\cite{Liu:1997km, Feng:1998hu, Chang:1999dai, Benatti:2000ni, Kalpdor:2000pas, Gago:2001mo, Benatti:2001fa, Fogli:2003li, Ashie:2004sk, Luis:2005rg, Barenboim:2006os, Farzan:2008zv, Oliveira:2010zzd, Oliveira:2013nua, Oliveira:2016asf, Coelho:2017byq, Carpio:2017nui, Carrasco:2018sca, Buoninfante:2020iyr, deHolanda:2019tuf, JUNO:2021wang, cheng:2022man, carrasco:2022dia} and it is different in its origin from the quantum mechanical wave packet decoherence~\cite{Giunti:1998kim, Blennow:2005ohl, Akhmedov:2012her, Chang:2016chu, Gouvea:2021rom}. In this paper, we consider only the former case. 

In the recent years, significant research efforts have been carried out in constraining the decoherence parameters. In ref.~\cite{Lisi:2000zt}, the authors have analysed Super-Kamiokande atmospheric neutrino experiment data and obtained bounds on the decoherence parameter $\Gamma < 3.5 \times 10^{-23}~GeV$ (90\% CL). In ref.~\cite{deOliveira:2013dia} the bounds on the decoherence parameters were obtained using MINOS data while in ref.~\cite{gomesalg:2023}, data from MINOS and T2K experiments has been analysed to update the bounds on the decoherence parameter by assuming power law dependency on the neutrino energy. The bounds obtained from the analysis of solar and KamLAND data have been presented in ref.~\cite{fogligl:2007}. In a further analysis, in ref.~\cite{Gomes:2016ixi} the authors included the recent KamLAND data. In a seminal paper ref.~\cite{Coelho:2017zes}, the authors pointed out that the $2\sigma$ tension between the T2K and NO$\nu$A data could be elevated by considering environmental decoherence of strength $\Gamma = (2.3 \pm 1.1) \times 10^{-23}~GeV$. The authors of ref.~\cite{Coloma:2018ice} have analyzed the IceCube/DeepCore data on atmospheric neutrinos and reported the bounds on the $\Gamma $ considering $n = 0,~\pm 1, ~\pm 2$. Recently, in ref.~\cite{Collab:2024icecube} the IceCube collaboration reported bounds on $\Gamma$ with atmospheric neutrinos for $n = 0, 1, 2, 3$. References~\cite{Carpio:2019mass, Gomes:2019for} present the sensitivity studies of DUNE to the decoherence parameters. 
The authors in ref.~\cite{Romeri:2023cgt} have updated the limits on $\Gamma$ by analysing reactor data from KamLAND, Daya Bay, RENO and accelerator data from T2K, NOvA, MINOS/MINOS+. Additionally, they have simulated the upcoming JUNO and DUNE facilities to study their sensitivity to $\Gamma$. In a more recent study ~\cite{Collab:2024essnusb}, the authors have obtained the sensitivity of the future long baseline ESSnuSB experiment to bound the decoherence parameter $\Gamma$ and further investigated the CPV sensitivity of the experiment in the presence of decoherence.


In an open quantum system framework the neutrino system interacts with the environment, leading to the loss of quantum coherence among the neutrino states. This decoherence effect manifests as a damping term $e^{-\Gamma L}$ in the neutrino oscillation probabilities, where $\Gamma$ is the decoherence parameter and L is the distance travelled by the neutrinos. $\Gamma$ can be parameterized using power-law dependency i.e., $\Gamma \propto E_\nu^n$ where $E_\nu$ is the neutrino energy and the value of $n$ depends on the origin of the decoherence model. In this context, without delving into the origin of the decoherence effect, we perform a phenomenological study of the effect of quantum neutrino decoherence in the Protvino to ORCA (P2O) experiment. P2O is an upcoming neutrino oscillation experiment with a substantial baseline of 2595 km~\cite{Akindinov:2019p2o, KM3Net:2016zxf}. This very large baseline of P2O experiment, makes it an ideal candidate to study the impact of environmentally induced decoherence on the neutrino oscillations. 

This paper is structured as follows. In section \ref{sec2} we present a concise overview of the density matrix formalism which is employed to derive the neutrino oscillation probabilities. We discuss the experimental and simulation details of our analysis in section \ref{sec3}. In section \ref{sec4}, we examine the impact of environmental decoherence on various oscillation channels (baseline 2595 km), taking into account the power law dependencies of decoherence on neutrino energy. In addition, we study the consequences of assuming the presence of decoherence 
in the true spectrum and assess the significance of P2O to determine the mass hierarchy and CP violation. Finally, we conclude our results in section \ref{sec5}.
\section{Theoretical formulation}\label{sec2}

In an open quantum system framework, neutrino sub system can interact very weakly with the stochastic environment. Keeping a phenomenological approach, we discuss the dissipative effects of the decoherence on neutrino oscillations in a model-independent way. To achieve this we consider density matrix formalism and represent the time evolution of the density matrix using the Lindblad master equation~\cite{Lindblad:1976g, GKS:1976vit}, 

\begin{equation}
    \frac{d\rho_m(t)}{dt} = -i\left [H,\rho_m(t) \right] + \mathcal{D} \left[ \rho_m (t) \right]~,
    \label{eq:LME}
\end{equation}

where $\rho_m$ is the density matrix of neutrino in the mass basis and H is the Hamiltonian of the neutrino system. The decoherence effect is introduced through a dissipative term $\mathcal{D}[\rho_m (t)]$. Parameterization of the decoherence matrix is performed by imposing mathematical properties of the density matrix such as complete positivity and preserving trace normalization (probability must be positive and conserved w.r.t time). Imposing complete positivity gives the Lindblad form of the dissipator as~\cite{GKS:1978g} 
\begin{equation}
\begin{aligned}
    \mathcal{D} \left[ \rho_m (t) \right] &= \frac{1}{2}\sum_{n = 1}^{N^2 - 1} \left\{[\mathcal{V}_n , \rho_m \mathcal{V}_n^\dagger] + [\mathcal{V}_n \rho_m , \mathcal{V}_n^\dagger]\right\}~,
    \end{aligned}
        \label{eq:D-term}
\end{equation}
\noindent

where $N$ is the dimension of the Hilbert space and $\mathcal{V}_n$ are the operators arising from the interaction of the subsystem with the environment. In addition, we impose an increase in von Neumann entropy $ S = - Tr(\rho_m \ln \rho_m) $~\cite{Banks:1984bsp,Benatti:1988nar} and conservation of average energy $Tr(\rho_m H)$ of the neutrino system. These conditions lead to $\mathcal{V}_n = \mathcal{V}_n^\dagger$~, $[\mathcal{V}_n , H] = 0$ and imply that $\mathcal{V}_n$ and $H$ are simultaneously diagonalisable. For a three-flavor neutrino system ($N = 3$), operators $\mathcal{V}_n$ ($n = 1~\text{to}~ 8$ ) are the linear combinations of the Gell-Mann matrices. Substituting for $\mathcal{V}_n$ in eq.~(\ref{eq:D-term}) and using the above mentioned conditions simplifies the dissipator term to

\begin{equation}
    \mathcal{D} \left[ \rho_m (t) \right] = \begin{pmatrix}
        0 & -\Gamma_{21} \rho_{12}(t) & -\Gamma_{31} \rho_{13}(t) \\
        -\Gamma_{21} \rho_{21}(t) & 0 & -\Gamma_{32} \rho_{23}(t) \\
        -\Gamma_{31} \rho_{31}(t) & -\Gamma_{32} \rho_{32}(t) & 0
    \end{pmatrix},
    \label{eq:newD}
\end{equation}
with 
\begin{equation}
    \Gamma_{jk} = \Gamma_{kj} = \frac{1}{2}\sum_{n = 1}^8 (d_{n,j} - d_{n,k})^2
    \label{eq:Gamma_ij}~,
\end{equation}
where, $d_{n,j}$, $d_{n,k}$ are the diagonal elements of $\mathcal{V}_n$ operator and $j,k$ take the values $1,2,3$.

Considering vacuum Hamiltonian in the mass basis $H = diag(0,~\Delta_{21},~\Delta_{31})$, where $\Delta_{jk} = \Delta m_{jk}^2/2E$, $\Delta m_{jk}^2$ being the neutrino mass-square differences ($m_j^2 - m_k^2$), one can obtain $\comm{H}{\rho_m(t)}$ to be
\begin{equation}
    \begin{aligned}
         \comm{H}{\rho_m(t)} = \begin{pmatrix}
            0 & -\rho_{12}(t) \Delta_{21} & -\rho_{13}(t) \Delta_{31} \\
            \rho_{21}(t) \Delta_{21} & 0 & -\rho_{23}(t) \Delta_{32} \\
            \rho_{31}(t) \Delta_{31} & \rho_{32}(t) \Delta_{32} & 0
        \end{pmatrix}~.
    \end{aligned}
    \label{eq:Hrho}
\end{equation}

The density matrix (mass basis) at time t is obtained by substituting eq.~(\ref{eq:newD}), eq.~(\ref{eq:Hrho}) in eq.~(\ref{eq:LME}) as~(see ref.~\cite{Coelho:2017byq} and appendix of ref. \cite{Gomes:2019for} for comprehensive overview)
\begin{equation}
    \rho_m(t) = \begin{pmatrix}
        \rho_{11}(0) & \rho_{12}(0) \exp-(\Gamma_{21} + i \Delta_{21})^* t & \rho_{13}(0) \exp-(\Gamma_{31} + i \Delta_{31})^* t \\
        \rho_{21}(0) \exp-(\Gamma_{21} + i \Delta_{21}) t & \rho_{22}(0) &  \rho_{23}(0) \exp-(\Gamma_{32} + i \Delta_{32})^* t \\
        \rho_{31}(0) \exp-(\Gamma_{31} + i \Delta_{31})t &  \rho_{32}(0) \exp-(\Gamma_{32} + i \Delta_{32})t & \rho_{33}(0)
    \end{pmatrix}~.
    \label{eq:Rho-m}
\end{equation}

Eq.~(\ref{eq:Rho-m}) can be converted into the flavour basis using the modified PMNS mixing matrix $\tilde{U}$, as we include the MSW effect. 
We consider the modified-mixing matrix ($\tilde{U}$) up to the 1st order approximation, from ref.~\cite{Denton:2018dmp} as
\begin{equation}
    \tilde{\rho}_{\alpha} = \tilde{U}~\tilde{\rho}_m~\tilde{U}^\dagger~.
    \label{eq:Rho-f}
\end{equation}
The neutrino transition probability from initial flavor '$\nu_\alpha$' to final flavor '$\nu_\beta$' in terms of density matrix is
\begin{equation}
   \begin{aligned}
        P_{\alpha \beta}(t) &= Tr[\tilde{\rho}_\alpha (t) \tilde{\rho}_\beta (0)]~.
   \end{aligned}
   \label{eq:P1}
\end{equation}
The explicit form of the transition probability assuming ultra-relativistic neutrinos ($t \approx L$) is given by~\cite{Gomes:2019for,Coloma:2018ice}
\begin{equation}
\begin{aligned}
        P_{\alpha \beta}(L) &= \delta_{\alpha \beta} - 2\sum_{j > k} Re \left( \tilde{U}_{\beta j} \tilde{U}_{\alpha j}^* \tilde{U}_{\alpha k} \tilde{U}_{\beta k}^* \right) + 2\sum_{j > k} Re \left( \tilde{U}_{\beta j} \tilde{U}_{\alpha j}^* \tilde{U}_{\alpha k} \tilde{U}_{\beta k}^* \right) \exp(-\Gamma_{jk} L) \cos(\frac{\tilde{\Delta}m_{jk}^2}{2E}L) \\& + 2\sum_{j > k} Im \left( \tilde{U}_{\beta j} \tilde{U}_{\alpha j}^* \tilde{U}_{\alpha k} \tilde{U}_{\beta k}^* \right) \exp(-\Gamma_{jk} L) \sin(\frac{\tilde{\Delta}m_{jk}^2}{2E}L)~.
        \end{aligned}
        \label{eq:Pab}
\end{equation}

Note that in the absence of the decoherence effect i.e., $\Gamma_{jk} = 0$ the probability expression reduces to the standard oscillation probability.

Let us examine eq.~(\ref{eq:Pab}) for the case of $\nu_e$ appearance probability term by term. For the case of $\alpha = \mu$, $\beta = e$ the non-zero terms of eq.~(\ref{eq:Pab}) take the form 

\begin{equation}
\begin{aligned}
        I &= - 2\sum_{j > k} Re \left( \tilde{U}_{e j} \tilde{U}_{\mu j}^* \tilde{U}_{\mu k} \tilde{U}_{e k}^* \right) ~.
        \end{aligned}
        \label{eq:Pmue-I}
\end{equation}

\begin{equation}
\begin{aligned}
        II &= 2\sum_{j > k} Re \left( \tilde{U}_{e j} \tilde{U}_{\mu j}^* \tilde{U}_{\mu k} \tilde{U}_{e k}^* \right) \exp(-\Gamma_{jk} L) \cos(\frac{\tilde{\Delta}m_{jk}^2}{2E}L) 
        \end{aligned}
        \label{eq:Pmue-II}
\end{equation}

\begin{equation}
\begin{aligned}
        III &= 2\sum_{j > k} Im \left( \tilde{U}_{e j} \tilde{U}_{\mu j}^* \tilde{U}_{\mu k} \tilde{U}_{e k}^* \right) \exp(-\Gamma_{jk} L) \sin(\frac{\tilde{\Delta}m_{jk}^2}{2E}L)~.
        \end{aligned}
        \label{eq:Pmue-III}
\end{equation}

\begin{itemize}
    \item The $term~I$ is not affected by the decoherence parameter $\Gamma_{jk}$. When $j=3$, $k=2$ this term presents a resonance in the appearance probability at energy $E_{\nu} \sim 10$ GeV. 
    
    \item In the $term~II$ (eq.~(\ref{eq:Pmue-II})) and $term~III$ (eq.~(\ref{eq:Pmue-III})), $\Gamma_{jk}$ appear in the dissipative form $\sim \exp(-\Gamma_{jk}L)$ where L is the distance travelled by the neutrino beam, $\Gamma_{jk}$ are the decoherence parameters that quantify the strength of the decoherence. Here, the coherence length can be obtained by 
    $L_{coh} = \frac{1}{\Gamma_{jk}}$.
    
    \item In the absence of the damping term $\exp(-\Gamma_{32}L)$, $terms~II$ and $term~III$ together cause a depression/dip in the probability and in turn cancel the resonance posed by $term~I$. However, in the presence of decoherence this dip disappears and we see the resonance around $E \sim 10$ GeV in the appearance probability. A detailed analysis regarding this interplay between $terms~I$, $II$ and $III$ can be found in ref.~\cite{Gomes:2019for}.
    
    \item To illustrate the above mentioned details regarding $\nu_e$-appearance probability, we plot $term~I$, $term~(II + III)$ (without decoherence), $term~(II + III)$ (with decoherence) for all values of $j,k = 1, 2, 3$ where $j>k$ in fig.~\ref{fig:prob-expl}.
    
    \item We obtain three sets of $jk = 32,31,21$. In the left, middle and right plot of fig.~\ref{fig:prob-expl}, we have shown the effect of $\exp(-\Gamma_{32}L)$, $\exp(-\Gamma_{31}L)$, $\exp(-\Gamma_{21}L)$, respectively. In each plot we have presented three curves, blue (dot-dashed), green (dashed) and magenta (dotted), corresponding to $term~I$, $term~(II + III)$ (without decoherence), $term~(II + III)$ (with decoherence). 
    
    \item In the left plot (where $jk = 32$), we observe that in absence of decoherence $term~I$ (bump) and $term~(II + III)$ (dip) cancel each other. However in the presence of decoherence, the dip in the   
    $term~(II + III)$ gets removed and the bump in $(term~I)$ causes a peak in the probability around E $\sim$ 10 GeV. 
    
    \item However, in the case of middle and right figures where the terms corresponding to $j=3$,~$k=1$ and $j=2$,~$k=1$ are plotted, we don't see any resonance or dip in the terms around $E \sim$ 10 GeV.

    \begin{figure*}[!htbp]
    \includegraphics[width=0.325\linewidth,height=5cm]{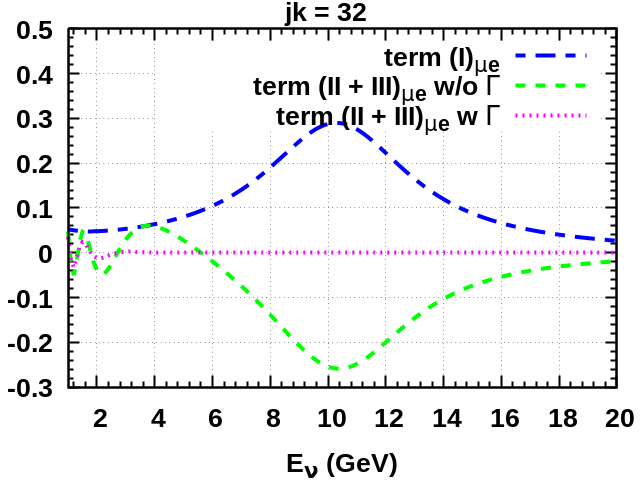}
    \includegraphics[width=0.325\linewidth,height=5cm]{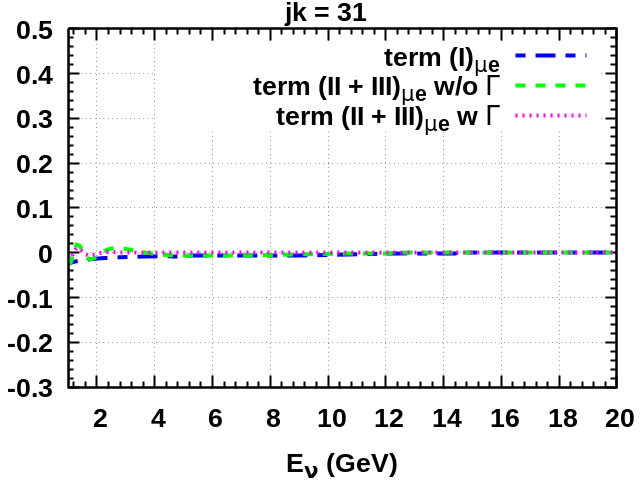}
    \includegraphics[width=0.325\linewidth,height=5cm]{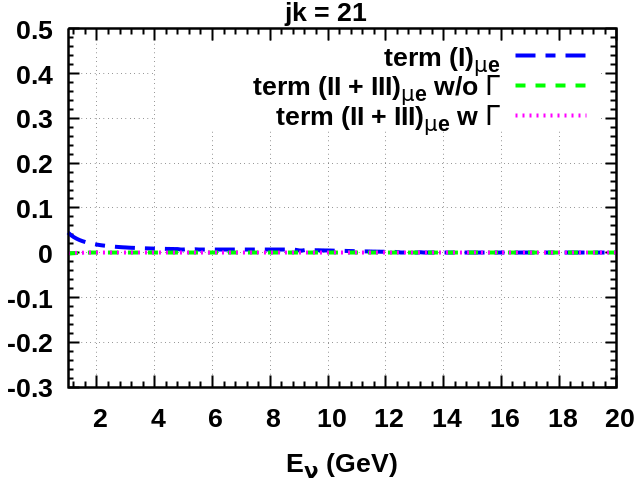}
    \caption{Illustrative plots for the elements term I, term II + III (without decoherence), term II + III (with decoherence) as function of $E_\nu$. The left, middle and right plots correspond to $jk = 32,31,21$ respectively.}
    \label{fig:prob-expl}
    \end{figure*}
    
    \item This explains why when a model assumes $\Gamma_{32} = 0$ (for instance in Case-II where $\Gamma_{21} = \Gamma_{31},~ \Gamma_{32} = 0$) the appearance probability doesn't get impacted by the presence of decoherence. 
\end{itemize}
We consider a general power law dependency of $\Gamma_{jk}$ on the neutrino energy given by 
\begin{equation}
    \Gamma_{jk}(E_\nu) = \Gamma_0\left(\frac{E_\nu}{E_0}\right)^n~,
    \label{G-powerlaw}
\end{equation}
where, $\Gamma_0$ is constant, $E_0$ is the reference energy taken as 1 GeV and $n = 0,\pm 1,\pm 2$. 
Different physical origins explain the decoherence phenomena leading to different integral power law dependencies~\cite{Romeri:2023cgt}. 

From eq.~(\ref{eq:Gamma_ij}) it is clear that the $\Gamma_{jk}$'s are not independent of each other. In our analysis, we consider the cases mentioned in table-\ref{table:1} where at least two $\Gamma_{jk}$'s are activated simultaneously. Each case has a unique decoherence matrix, eq.~(\ref{eq:newD}), and thus damps different neutrino oscillation channels differently. 
\begin{table}[!htbp]\centering
\begin{tabular}{ |c|c| } 
 \hline\hline
 Cases & Assumptions on $\Gamma_{jk}$ \\
 \hline
 Case-I & $\Gamma_{21} = \Gamma_{31} = \Gamma_{32} \neq 0$ \\ 
 Case-II & $\Gamma_{21} = \Gamma_{31},~ \Gamma_{32} = 0 $ \\ 
 Case-III & $\Gamma_{21} = \Gamma_{32},~ \Gamma_{31} = 0$ \\ 
 Case-IV & $\Gamma_{31} = \Gamma_{32},~ \Gamma_{21} = 0$  \\ 
 \hline\hline
\end{tabular}
\captionsetup{justification=centering}
 \caption{The decoherence models considered in this work.}
\label{table:1}
\end{table}
In this work, we study the effect of the different power law dependencies given in eq.~(\ref{G-powerlaw}), for each of the cases in table-\ref{table:1}.

\section{Experimental and Simulation details}\label{sec3}

Protvino to ORCA (P2O) is an upcoming long baseline neutrino oscillation experiment. The neutrino beam travels for about 2595 km from Protvino accelerator complex before it reaches the 8 MT water cherenkov KM3NeT/ORCA detector. The initial plan is to have a 90 kW proton beam resulting in $0.8 \times 10^{20}$ protons on target (POT) per year. A later upgrade of the beam facility targets a 450 kW proton beam. The resultant muon neutrino beam of $2 - 7$ GeV passes through the upper mantle of the earth with matter density $\sim ~3.3~gm/cm^3$ before it reaches the far detector that is currently being constructed in Mediterranean Sea. The oscillation channels of utmost interest are $\nu_\mu \rightarrow \nu_e,~ \nu_\mu \rightarrow \nu_\mu$ channels and the corresponding anti-neutrinos channels.

We assumed a total exposure of 6 years where 3 years run-time is for the neutrino beam and 3 years for the anti-neutrino beam. In the standard three flavor oscillation scenario the simulated data  from 3 years of neutrino mode (assuming 90 kW beam) resulted in a total of $\sim 4800~\nu_{\mu}$ disappearance events and $\sim 2600~\nu_e$ appearance events at the far detector. On the other hand, in the anti-neutrino mode we found $\sim $ 1406 disappearance events and $\sim $ 80 appearance events assuming normal ordering, $\theta_{23} = 45\degree$ and $\delta_{CP} = 270\degree$. The fraction of events classified as tracks (muon) and showers (electron) are taken from fig.~99 of ref.~\cite{KM3Net:2016zxf}. The energy resolution of the detector is considered $\approx$ 30\% as given in ref.~\cite{Akindinov:2019p2o}.
Under the systematic uncertainties, we considered a $5\%$ on the signal normalisation, $12\%$ on the background normalisation and a $11\%$ tilt error on both signal and background. 

We have used the GLoBES software~\cite{Huber:2004ka, Huber:2007ji} to simulate the P2O experiment and perform our analysis. Considering the standard three flavour oscillation picture we reproduced the event spectra w.r.t the true neutrino energy that is reported in ref.~\cite{Akindinov:2019p2o}. In this work, we simulated the data corresponding to 90 kW proton beam resulting in $0.8 \times 10^{20}$
POT/year and also 450 kW beam corresponding to the $4.0 \times 10^{20}$ POT/year and total run time 6 years (3 years for $\nu$ and 3 years for $\bar{\nu}$ mode) in each case. 
\begin{table}[!htbp]\centering
\begin{tabular}{ |c|c|c| } 
 \hline\hline
 Parameters & True values & $3\sigma$ ranges \\[0.5ex]
 \hline
 $\sin^2{\theta_{12}}$ & $0.304$ & Fixed \\ [0.5ex]
 $\sin^2{\theta_{13}}$ & $0.0222$ & Fixed \\ [0.5ex]
 $\sin^2{\theta_{23}}$ & $0.573$ & $[0.405 : 0.620]$ \\ [0.5ex]
 $\delta_{CP}$ & $194\degree$ & $[0 : 360\degree]$ \\ [0.5ex]
 $\frac{\Delta m^2_{21}}{10^{-5}~eV^2}$ & $7.42$ & Fixed \\ [0.5ex]
 $\frac{\Delta m^2_{31}}{10^{-3}~eV^2}$ (NH) & $2.515$ & $[2.431 : 2.599]$ \\ [0.5ex]
 $\frac{\Delta m^2_{31}}{10^{-3}~eV^2}$ (IH) & $-2.498$ & $[-2.584 : -2.413]$\\[0.5ex]
 \hline\hline
\end{tabular}
\captionsetup{justification=centering}
 \caption{True oscillation parameters and $3\sigma$ ranges have been considered in our analysis are taken from NuFIT~\cite{Esteban:2020nufit}.}
 \label{table:2}
\end{table}
Later we incorporated a new oscillation probability engine into GLoBES by taking into account the effect of decoherence on neutrino propagation.
The best-fit values of the standard oscillation parameters and their $3\sigma$ ranges used in the analysis are given in table-\ref{table:2}. For the statistical analysis, the $\chi^2$ function is calculated using Poisson chi-square function given by

\begin{equation}
    \chi^2=\min _{\alpha_s,\alpha_b} \sum_{\text {channels }} 2 \sum_i\left[N_i^{\mathrm{test}} - N_{i}^{\mathrm{true}}+N_{i}^{\mathrm{true}} \log \left(\frac{N_{i}^{\mathrm{true}}}{N_{i}^{\mathrm{test}}}\right)\right] + \alpha_s^2 + \alpha_b^2~,
    \label{eq:chi-sq}
\end{equation} 
where, $N_i^{\mathrm{test}}$ and $N_i^{\mathrm{true}}$ are the number of test and true events (signal + background) in $i$-th bin, respectively. $\alpha_s$ and $\alpha_b$ are the signal and background normalisation errors treated using the $\emph{pull method}$~\cite{Fogli:2002pt, Huber:2002mx}. The analysis window of the neutrino energy is considered from 2 to 12 GeV. We marginalize over the $\delta_{CP}$, $\theta_{23}$, $\Delta m_{31}^2$ in their $3\sigma$ ranges without assuming any priors. 
\section{Results and discussion}\label{sec4}

In this segment, we outline the findings of our analysis across four subsections, each focusing on the cases mentioned in table-\ref{table:1}. The sensitivity studies have been performed for both 90 kW and 450 kW proton beams in all the subsections and the corresponding bounds of $\Gamma_{jk}$ have been reported in table~\ref{table:3} and table~\ref{table:4} respectively. Henceforth, we will refer to the simulations with 90 kW proton beam as P2O and that with 450 kW beam as P2O-upgrade.
 
In each subsection, firstly we show the effect of non-zero decoherence on the three oscillation channels $\nu_\mu \rightarrow \nu_e$ (left), $\nu_\mu \rightarrow \nu_\mu$ (middle) and $\nu_\mu \rightarrow \nu_\tau$ (right) for P2O baseline. To achieve this, we have considered the true values of the standard oscillation parameters given in the second column of table~\ref{table:2} allowing NH. However, for the sensitivity studies we have taken into account only the relevant oscillation channels $\nu_\mu \rightarrow \nu_e$, $\nu_\mu \rightarrow \nu_\mu$, $\bar{\nu}_\mu \rightarrow \bar{\nu}_e$, $\bar{\nu}_\mu \rightarrow \bar{\nu}_\mu$~. Each figure contains the standard neutrino propagation in matter (SM) and the non-zero decoherence assuming energy dependency index in eq.~(\ref{G-powerlaw}) as $n = -2,~-1,~0,~1,~2$. In each plot, we show five curves where green, magenta, blue, dark green and red lines represent $n = -2,~-1,~0,~1~\text{and}~2$ respectively.

Secondly, to quantify the sensitivity of P2O experiment to the decoherence parameter $\Gamma_{jk}$, we plot $\chi^2$ as a function of $\Gamma_{jk}$(test). We simulate the data by taking $\Gamma_{jk} = 0$(true) and the test spectrum by considering $\Gamma_{jk} \neq 0$. We calculate $\chi^2$ as described in sec~\ref{sec3}, and obtain
\begin{equation}\label{eq:chi-Gamma}
    \Delta\chi_{\Gamma}^2 = \chi^2 (\Gamma (true) = 0, \Gamma (test) \neq 0)~,
\end{equation}
after marginalizing over $\theta_{23},~\delta_{CP}$ and $(\Delta m_{31}^2)_{NH}$ in the test spectrum. Further, we provide the expected bounds on $\Gamma_{jk}$ obtained for P2O and P2O-upgrade in table~\ref{table:3} and table~\ref{table:4} respectively.

Later assuming the presence of decoherence in the true spectrum we study the effect of non-zero $\Gamma_{jk}$ on the determination of mass hierarchy and the discovery CP violation at P2O experiment. This study shows the consequence of not assuming decoherence in the theoretical hypothesis while it is present in nature. For this, we consider the $3\sigma$ value of $\Gamma_{jk}$ obtained from the table~\ref{table:3} and table~\ref{table:4} as the true values in each subsection.


The $\Delta \chi_{MH}^2$ is defined as
\noindent
\begin{equation}\label{eq:chi-MH}
    \Delta \chi_{MH}^2 = \chi_{true}^2 (\Gamma \neq 0, \Delta m_{31}^2 > 0) - \chi_{test}^2 (\Gamma = 0, \Delta m_{31}^2 < 0)~.
\end{equation}
Here, we marginalize over the test parameters $\theta_{23}$ and $\delta_{CP}$. 

The $\Delta \chi_{CPV}^2$ is given by

\begin{equation}\label{eq:chi-CPV}
    \begin{aligned}
        &\Delta \chi_0^2 = \chi_{true}^2 (\delta_{CP}(true),\Gamma \neq 0) - \chi_{test}^2 (\delta_{CP} = 0,\Gamma = 0)~,\\
        & \Delta \chi_\pi^2 = \chi_{true}^2 (\delta_{CP}(true),\Gamma \neq 0) - \chi_{test}^2 (\delta_{CP} = \pi, \Gamma = 0)~,\\
        & \Delta \chi_{CPV}^2 = min [\Delta\chi_0^2, \Delta\chi_\pi^2]~.
    \end{aligned}
\end{equation}
We marginalize over the test parameters $\theta_{23}$ and $(\Delta m_{31}^2)_{NH}$. 
The significance ($\sigma$) of P2O to discover the MH and CPV is obtained by using $\sigma_{MH/CPV} = \sqrt{\Delta\chi_{MH/CPV}^2}$~. 

\subsection{{Case-I: $\Gamma_{21} = \Gamma_{31} = \Gamma_{32} \neq 0$}\label{sub:case1}}
\begin{figure*}[!t]
\includegraphics[width=0.325\linewidth]{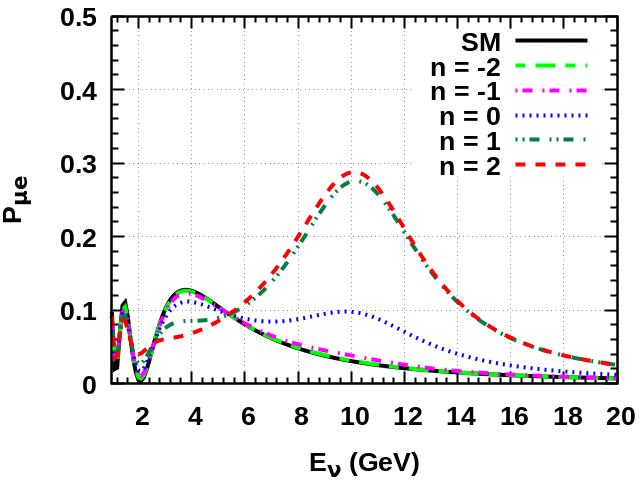}
\includegraphics[width=0.325\linewidth]{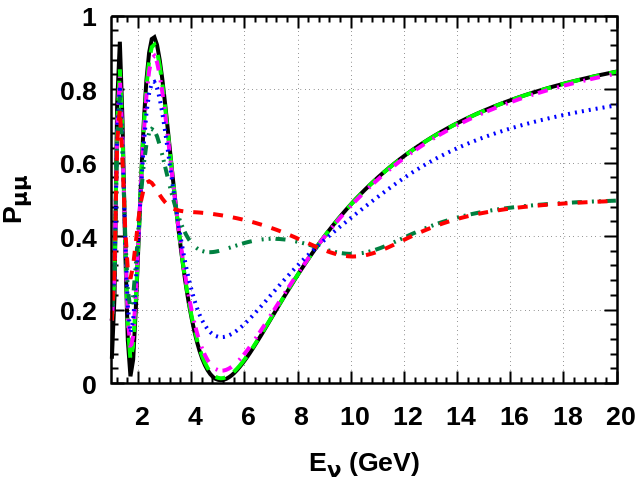}
\includegraphics[width=0.325\linewidth]{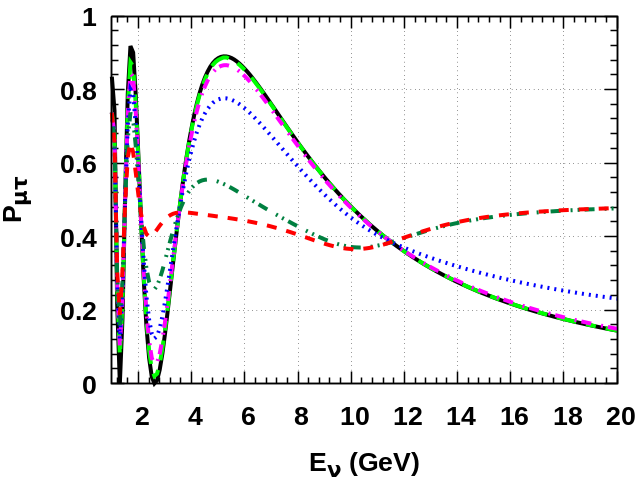}
\caption{Probability versus neutrino energy for $\nu_e$ appearance (left), $\nu_\mu$ disappearance (middle), $\nu_\tau$ appearance (right) considering case-I ( $\Gamma_{21} = \Gamma_{31} = \Gamma_{32} = 2.3 \times 10^{-23} $ GeV). The different colors correspond to the probabilities for the standard case and various $n$ dependencies are as mentioned in the legend.}
\label{fig1}
\end{figure*}

In this case, we assume $\Gamma_{21}$, $\Gamma_{31}$ and $\Gamma_{32}$ to be equal to $2.3 \times 10^{-23}$ GeV and non-zero. Fig.~(\ref{fig1}) shows the probability of the three oscillation channels, $\nu_\mu \rightarrow \nu_e$, $\nu_\mu \rightarrow \nu_\mu$ and $\nu_\mu \rightarrow   \nu_\tau$ in the left, middle and right panels, respectively. Legends corresponding to each plot are mentioned in the figure. It can be seen that all the three oscillation probabilities are significantly modified for the positive powers of $n \geq 0$ when compared to the SM case (black curve). Additionally, in the left plot, when $n \geq 0$, there is an extra peak around 10 GeV in the $P_{\mu e}$ channel. The probability $P_{\mu\mu}$ ($P_{\mu\tau}$) is higher (lower) than the SM probability in the energy window $E_\nu \sim 4-9 ~GeV$ and lower (higher) in the ranges $E_\nu \sim 2-4 ~GeV$ and $E_\nu > 9~GeV$. The first oscillation peak (dip) in $\nu_e$ appearance ( $\nu_{\mu}$ disappearance) channel is effected by the decoherence for the chosen set of decoherence parameters $\mathcal{O}(10^{-23} GeV)$. However, we noticed that this effect is not significant when the reference value of decoherence parameters is chosen to be $\mathcal{O}(10^{-24} GeV)$ and the extra peak around 10 GeV remains in place with lesser amplitude.



\begin{figure*}[!t]
\includegraphics[width=0.325\linewidth]{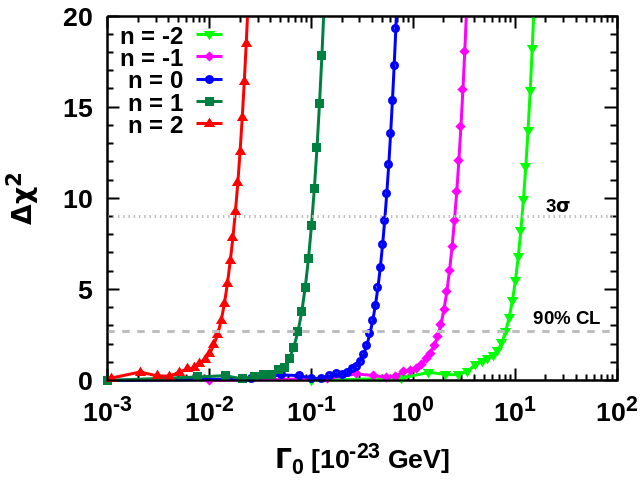}
\includegraphics[width=0.325\linewidth]{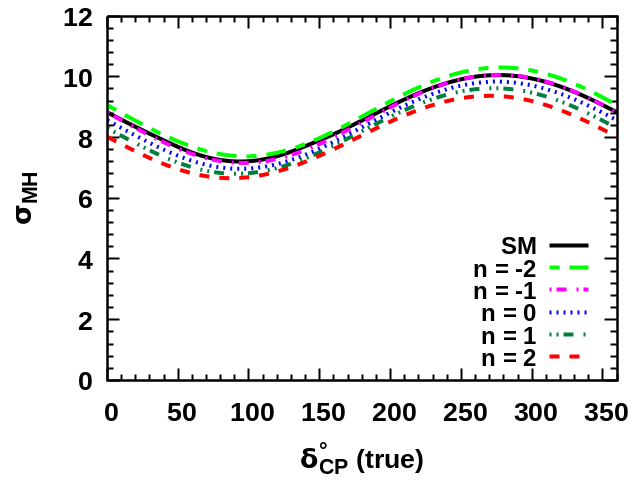}
\includegraphics[width=0.325\linewidth]{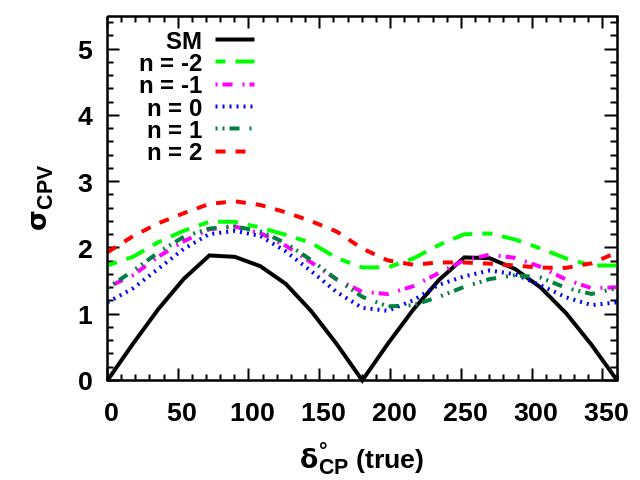}
\caption{Case-I for P2O experiment (90kW beam). In the left, we present $\Delta \chi^2$ as a function of $\Gamma_0$ (test) based on power law index $n = 0, \pm 1, \pm 2$. Horizontal dashed and dotted lines stands for 90\% and $3\sigma$ CL respectively. In the middle and right panels we show the significance of the experiment to discover MH and CP violation.}
\label{fig2}
\end{figure*}
\begin{figure*}[!t]
\includegraphics[width=0.325\linewidth]{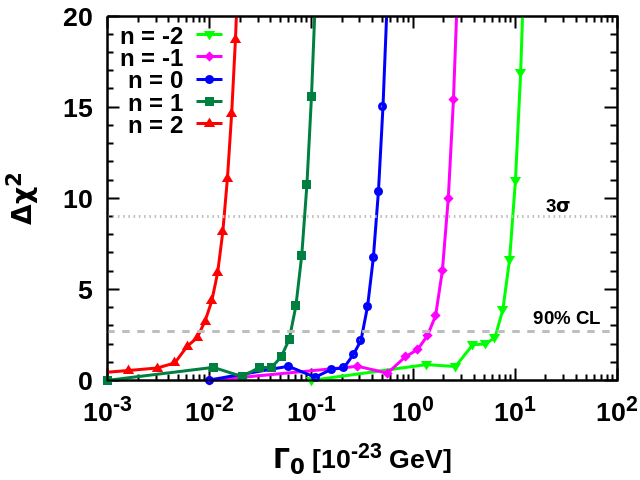}
\includegraphics[width=0.325\linewidth]{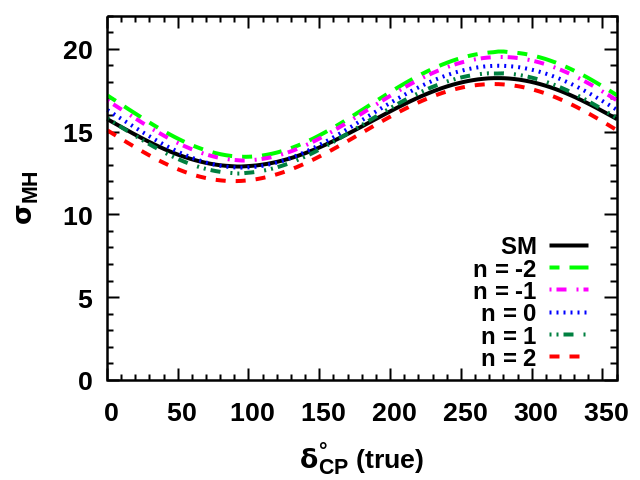}
\includegraphics[width=0.325\linewidth]{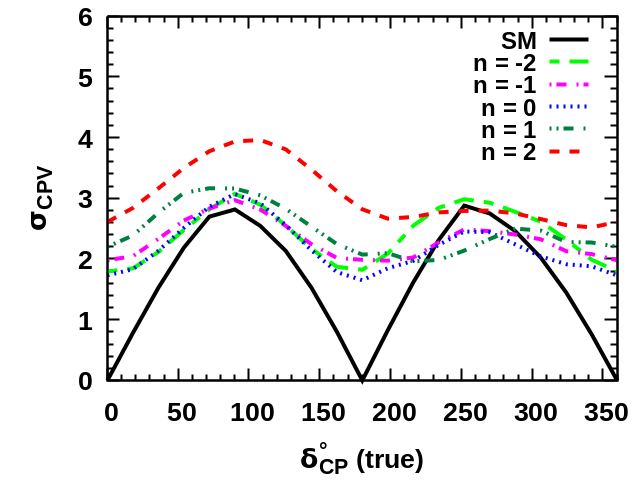}
\caption{Case-I for P2O-upgrade (450 kW beam power). In the left, we present $\Delta \chi^2$ as a function of $\Gamma_0$ (test) based on power law index $n = 0, \pm 1, \pm 2$. Horizontal dashed and dotted lines stands for 90\% and $3\sigma$ CL respectively. In the middle and right panels we show the significance of the experiment to discover MH and CP violation.}
\label{fig3}
\end{figure*}


In the left plots of fig.~(\ref{fig2}) and fig.~(\ref{fig3}), we constrain $\Gamma_{jk}$ presuming the conditions on decoherence parameters as given in case I. We calculate the test statistics using eq.~(\ref{eq:chi-Gamma}). The horizontal dashed and dotted lines correspond to $90\%$ ($\sim 1.64\sigma$) and $3\sigma$ confidence level (CL) respectively. 
The corresponding bounds of these parameters for different $n$ values are tabulated in the first column of table-\ref{table:3} and  table-\ref{table:4} respectively. From left most plots of fig.~(\ref{fig2}) and fig.~(\ref{fig3}), it can be seen that the sensitivity to the decoherence parameter has slightly improved after the proton beam upgrade to 450 kW. It can be verified from the first column of table-\ref{table:3} and -\ref{table:4}, that the $3\sigma$ bounds on $\Gamma_{jk}$ have improved by one order of magnitude for the cases with $n \geq 0$ for the P2O-upgrade (450 kW). 

The $\Delta \chi_{MH}^2$ vs $\delta_{CP}$ (true) and $\Delta \chi_{CPV}^2$ vs $\delta_{CP}$ (true) are calculated at $3\sigma$ value of $\Gamma_{jk}$ using eq.~\ref{eq:chi-MH} and eq.~\ref{eq:chi-CPV}. The respective significance $\sigma_{MH}$ (middle panel) and $\sigma_{CPV}$ (right panel) are plotted for the standard case and for non-zero decoherence (with $n = 0, \pm 1, \pm 2$) in fig.(\ref{fig2}) and fig.(\ref{fig3}) respectively. For the 90 kW beam, we can see the MH sensitivity (middle plot) for $n = -2$ is higher than that for $n = 2$. The significance $\sigma_{MH}$ is maximum around $\delta_{CP} \text{(true)} = 270\degree$. In the case of 450 kW beam, the over all significance to the mass hierarchy is higher ($\sim 12 \sigma$) and the order of decoherence curves corresponding to different values of $n$ obey a similar trend to that in 90 kW beam. However, since the MH sensitivity of P2O (for both 90 kW and 450 kW) is very high, one can note that the effect of decoherence is not prominent on the determination of MH. The same conclusions have been observed in the plots where true hierarchy is assumed to be IH. 

In the right most plots of fig.~(\ref{fig2}) and fig.~(\ref{fig3}), we observe that there is non-zero significance corresponding to CP conserving values of $\delta_{CP} = 0, \pm \pi$ for all values of $n$. This indicates that the new physics phenomenon of decoherence could induce an extrinsic (fake) CP phase and mislead the discovery of the CP violation at P2O.

\subsection{\texorpdfstring{Case-II (Solar limit I) : $\Gamma_{21} = \Gamma_{31},~ \Gamma_{32} = 0$}{TEXT}{}}\label{sub:case2}

\begin{figure*}[!t]
\includegraphics[width=0.325\linewidth]{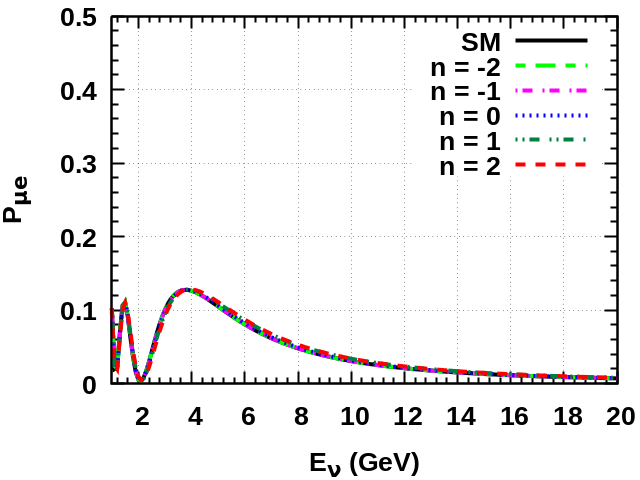}
\includegraphics[width=0.325\linewidth]{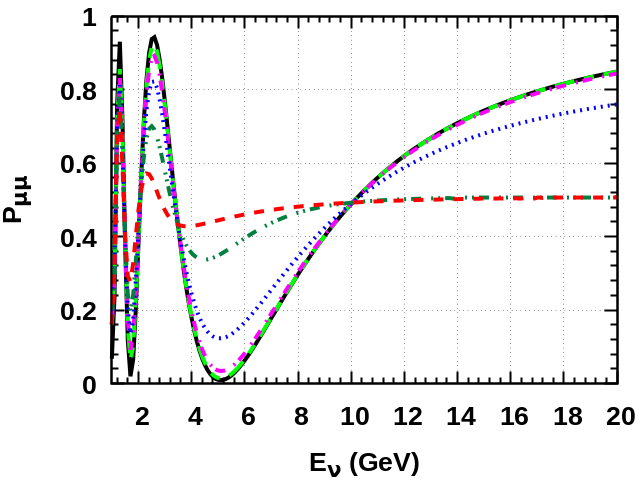}
\includegraphics[width=0.325\linewidth]{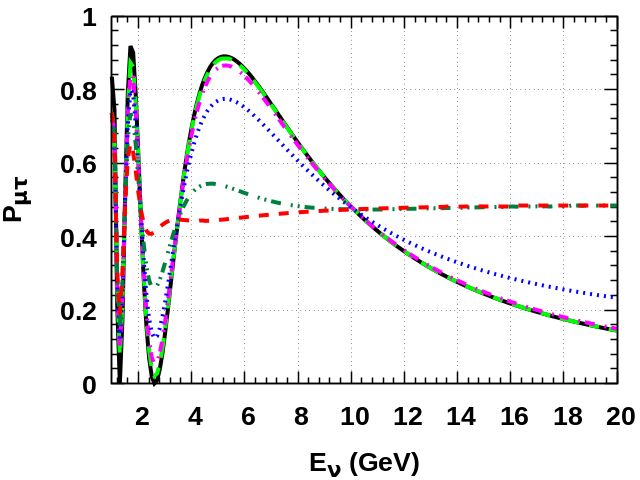}
\caption{Probability versus neutrino energy for $\nu_e$ appearance (left), $\nu_\mu$ disappearance (middle), $\nu_\tau$ appearance (right) considering case-II, $\Gamma_{21} = \Gamma_{31} = 2.3 \times 10^{-23}~ GeV,~ \Gamma_{32} = 0$. The different colors correspond to the probabilities for the standard case and various $n$ dependencies are as mentioned in the legend.}
\label{fig4}
\end{figure*}

The transition probabilities in this case are shown in fig.~(\ref{fig4}). From the left plot, one can see no notable change in the $\nu_e$-appearance probability compared to the SM for the chosen set of decoherence parameters and also no extra peak is observed at $E_\nu \sim 10~GeV$. In the middle plot, the $\nu_{\mu}$ disappearance probabilities are presented for different $n$. Clearly, the probability values significantly vary for the cases with $n \geq 0$. The same can be noted from the right most plot showing $\nu_\mu \rightarrow \nu_\tau$ probability. 

\begin{figure*}[!t]
\includegraphics[width=0.325\linewidth]{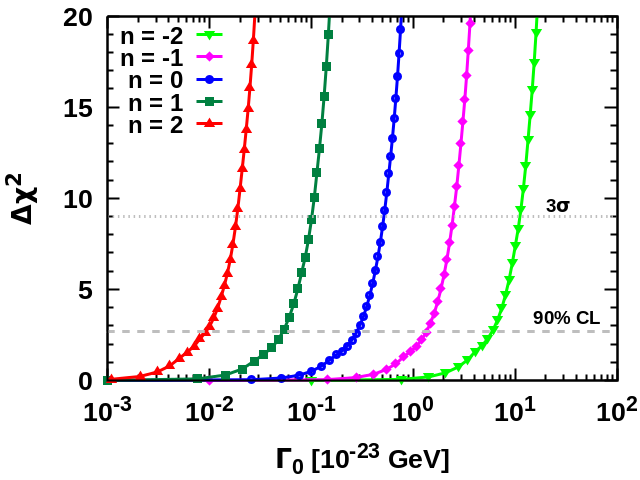}
\includegraphics[width=0.325\linewidth]{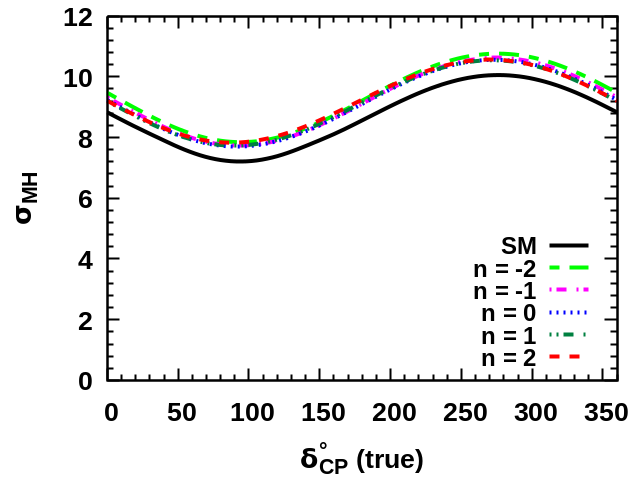}
\includegraphics[width=0.325\linewidth]{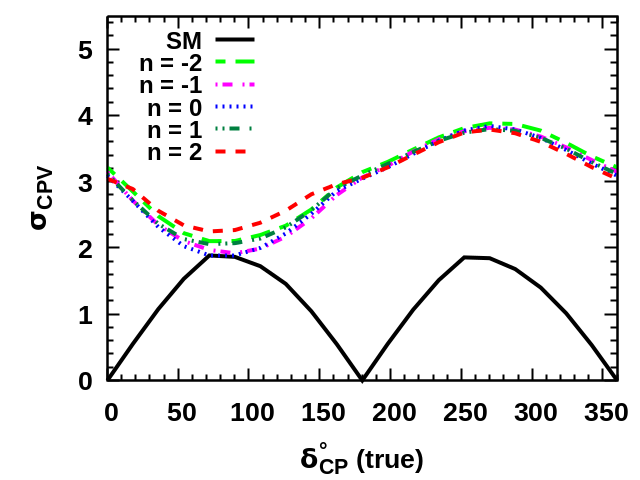}
\caption{Case-II for P2O experiment (90kW beam). In the left, we present $\Delta \chi^2$ as a function of $\Gamma_0$ (test) based on power law index $n = 0, \pm 1, \pm 2$. Horizontal dashed and dotted lines stands for 90\% and $3\sigma$ CL respectively. In the middle and right panels we show the significance of the experiment to discover MH and CP violation.}
\label{fig5}
\end{figure*}
\begin{figure*}[!t]
\includegraphics[width=0.325\linewidth]{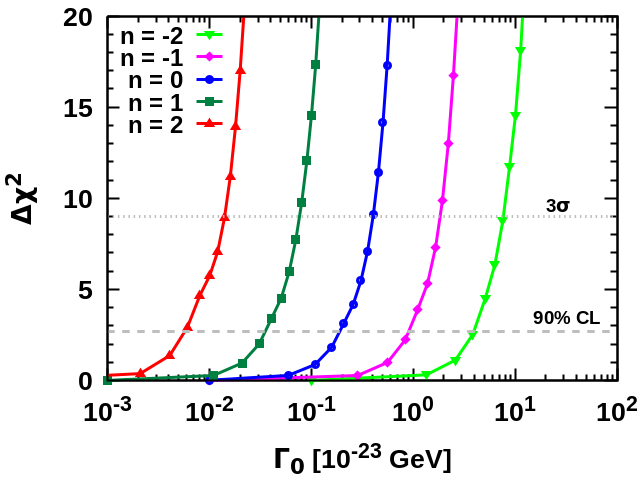}
\includegraphics[width=0.325\linewidth]{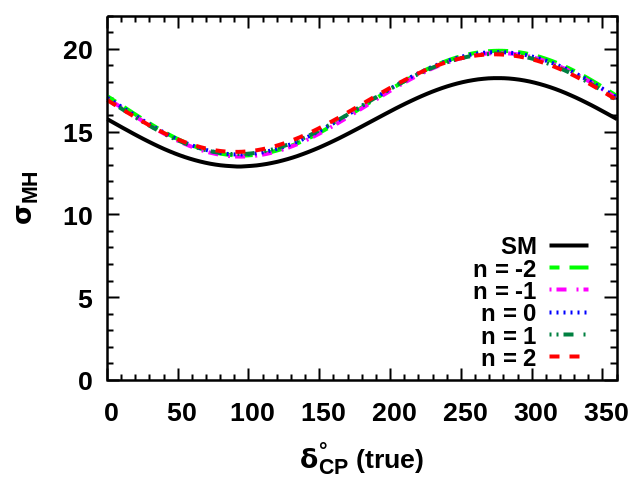}
\includegraphics[width=0.325\linewidth]{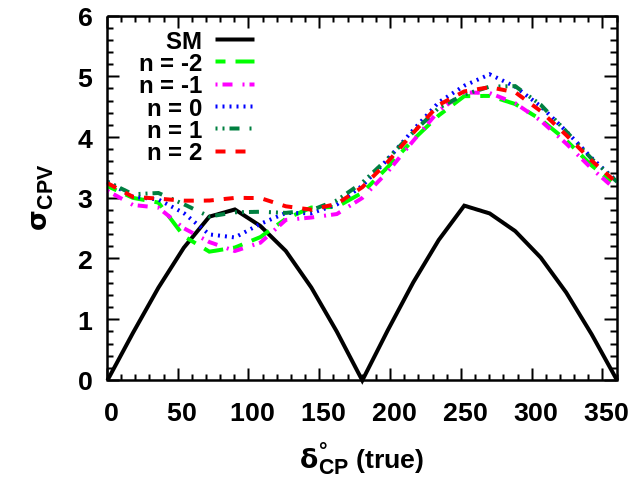}
\caption{Case-II for P2O-upgrade (450 kW beam power). In the left, we present $\Delta \chi^2$ as a function of $\Gamma_0$ (test) based on power law index $n = 0, \pm 1, \pm 2$. Horizontal dashed and dotted lines stands for 90\% and $3\sigma$ CL respectively. In the middle and right panels we show the significance of the experiment to discover MH and CP violation.}
\label{fig6}
\end{figure*}

To obtain the bounds on the decoherence parameters, we plot $\chi^2$ vs $\Gamma$ in the left panel of fig.~(\ref{fig5}) and fig.~(\ref{fig6}). The corresponding bounds of $90\%$ and $3\sigma$ CL on $\Gamma$ are listed in the second column of table-\ref{table:3} and table-\ref{table:4}. 
In the middle and the right figures, we present the significance $\sigma_{MH}$ and $\sigma_{CPV}$ with respect to true $\delta_{CP}$ using $3\sigma$ bounds obtained from the left plot. From the middle panel it is clear that $\sigma_{MH}$ has not altered significantly for all values of n. However, in the case of $\sigma_{CPV}$, we observe that for all true values of $\delta_{CP}$ the significance of the P2O experiment is significantly mislead due to the mismatch in the true spectrum and the fit hypothesis.

\subsection{\texorpdfstring{Case-III (Solar limit II) : $\Gamma_{21} = \Gamma_{32},~ \Gamma_{31} = 0$}{TEXT}{}}\label{sub:case3}

\FloatBarrier
\begin{figure*}[hbt!]
\includegraphics[width=0.325\linewidth]{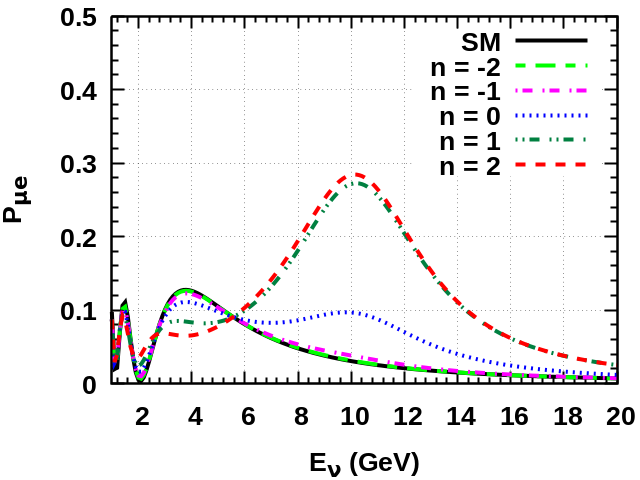}
\includegraphics[width=0.325\linewidth]{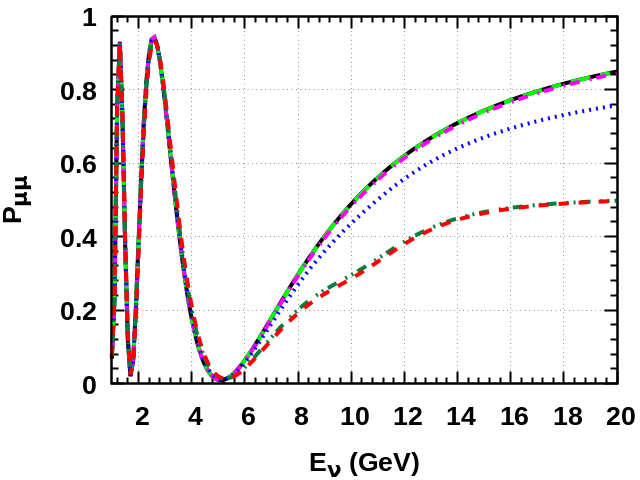}
\includegraphics[width=0.325\linewidth]{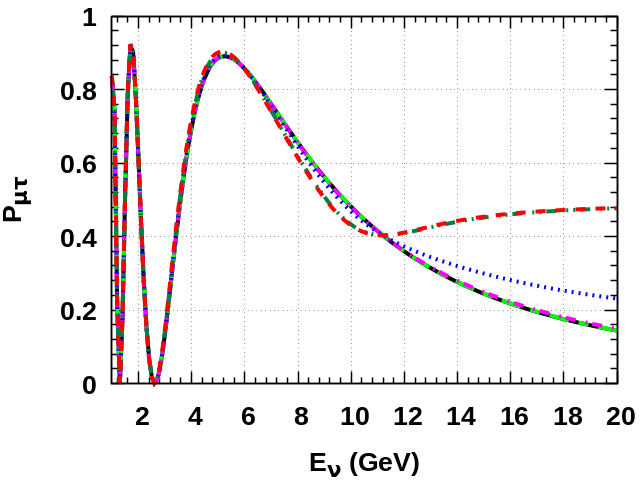}
\caption{Probability versus neutrino energy for $\nu_e$ appearance (left), $\nu_\mu$ disappearance (middle), $\nu_\tau$ appearance (right) considering case-III, $\Gamma_{21} = \Gamma_{32} = 2.3 \times 10^{-23}~ GeV,~ \Gamma_{31} = 0$}
\label{fig7}
\end{figure*}
\begin{figure*}[hbt!]
\includegraphics[width=0.325\linewidth]{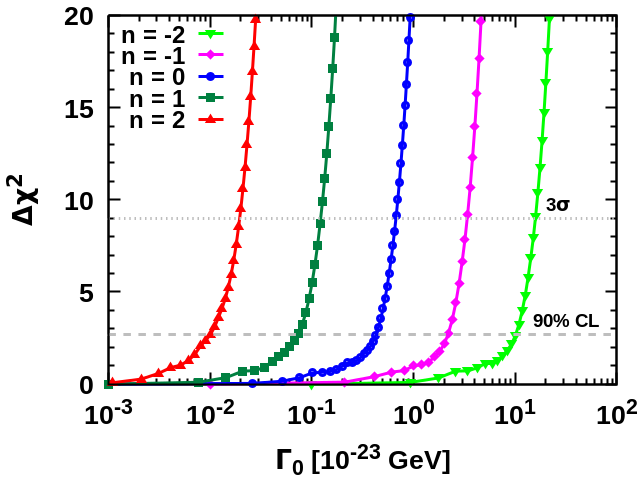}
\includegraphics[width=0.325\linewidth]{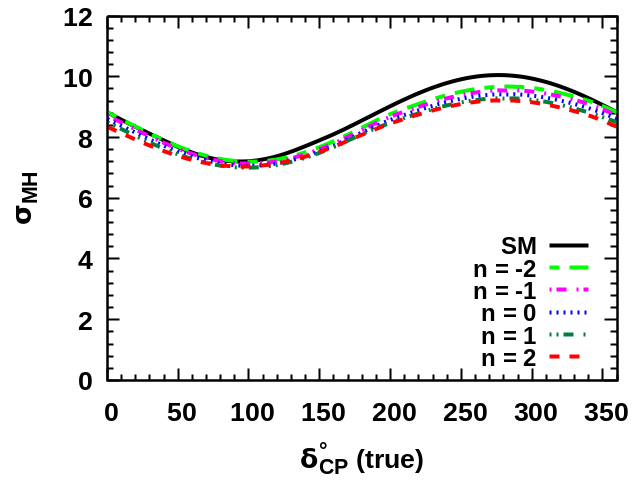}
\includegraphics[width=0.325\linewidth]{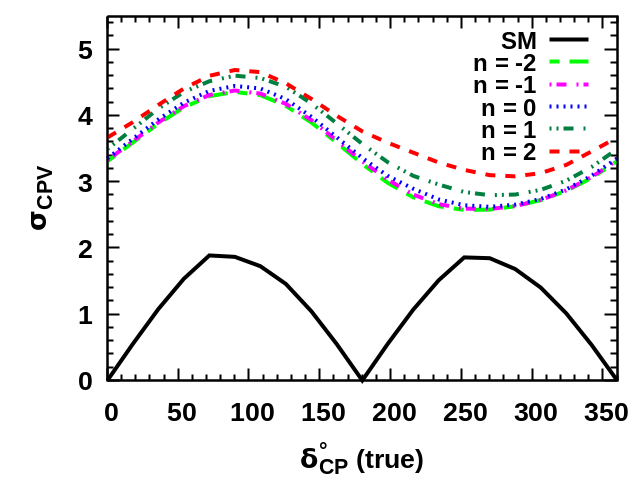}
\caption{Case-III for P2O experiment (90kW beam). In the left, we present $\Delta \chi^2$ as a function of $\Gamma_0$ (test) based on power law index $n = 0, \pm 1, \pm 2$. Horizontal dashed and dotted lines stands for 90\% and $3\sigma$ CL respectively. In the middle and right panels we show the significance of the experiment to discover MH and CP violation.}
\label{fig8}
\end{figure*}
\begin{figure*}[hbt!]
\includegraphics[width=0.325\linewidth]{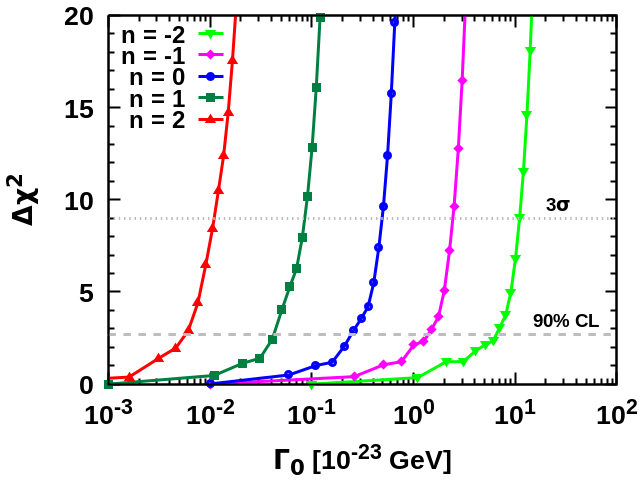}
\includegraphics[width=0.325\linewidth]{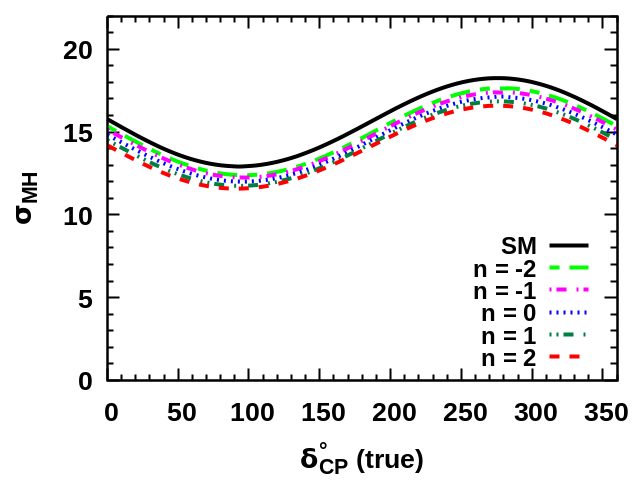}
\includegraphics[width=0.325\linewidth]{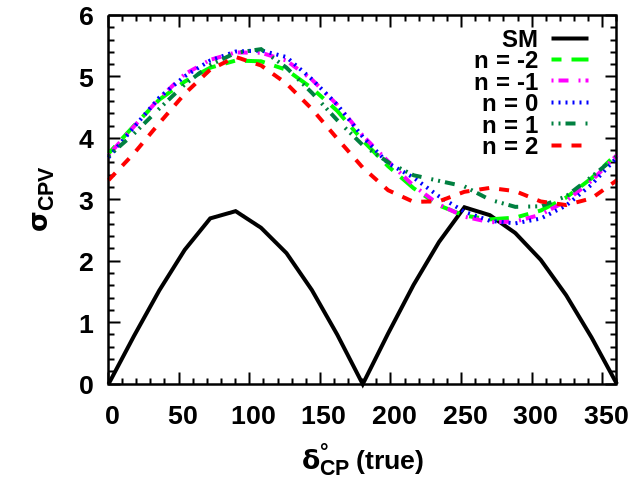}
\caption{Case-III for P2O-upgrade (450 kW beam power). In the left, we present $\Delta \chi^2$ as a function of $\Gamma_0$ (test) based on power law index $n = 0, \pm 1, \pm 2$. Horizontal dashed and dotted lines stands for 90\% and $3\sigma$ CL respectively. In the middle and right panels we show the significance of the experiment to discover MH and CP violation.}
\label{fig9}
\end{figure*}
We plot $\nu_\mu \rightarrow \nu_e$ transition in the left, $\nu_{\mu} \rightarrow \nu_{\mu}$ in the middle and $\nu_\mu \rightarrow \nu_\tau$ in the right considering case-III in fig.~(\ref{fig7}). Looking at the left plot one can state the $\nu_e$-appearance probability varies similar to case-I. The $\nu_\mu$-disappearance probability decreases for $E_\nu \geq 6~GeV$ and $\nu_\tau$-appearance probability increases for $E_\nu \geq 11~GeV$.

The bounds on the decoherence parameters in this case are shown in the left plots of fig.~(\ref{fig8}) and fig.~(\ref{fig9}) and are tabulated in the third column of table-\ref{table:3} and table-\ref{table:4}. In the middle panel, we plot the MH sensitivities where we can see the $\sigma$ for all $n$ are lesser than the values corresponding to the SM and the maximum ($\sim 20\sigma$) is visible around $270\degree$. 
However in the case of $\sigma_{CPV}$ for P2O and P2O-upgrade, the discovery potential to CP violation is wronged due to the presence of decoherence.
\subsection{\texorpdfstring{Case-IV (Atmospheric limit) : $\Gamma_{31} = \Gamma_{32},~ \Gamma_{21} = 0$}{TEXT}{}}\label{sub:case4}
\begin{figure*}[hbt!]
\includegraphics[width=0.325\linewidth]{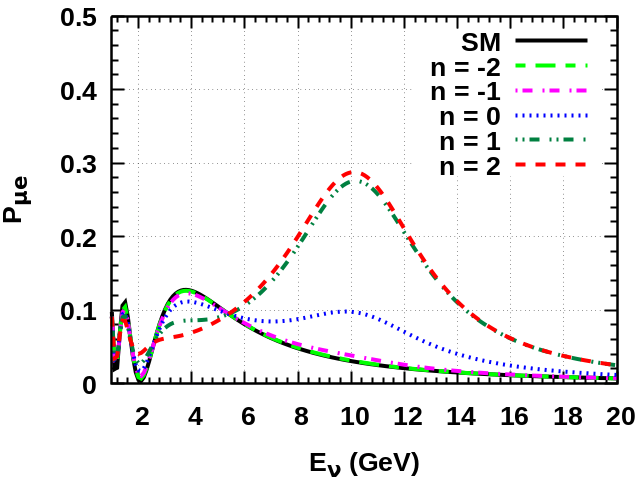}
\includegraphics[width=0.325\linewidth]{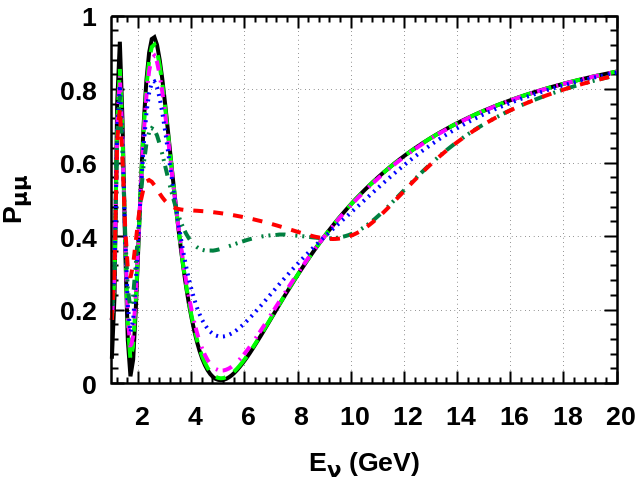}
\includegraphics[width=0.325\linewidth]{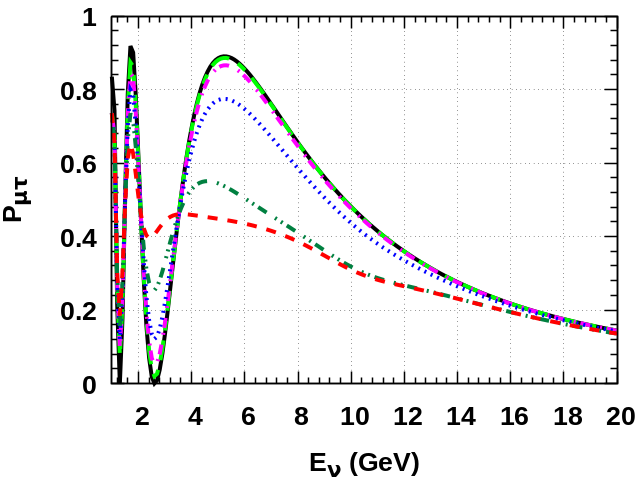}
\caption{Probability versus neutrino energy for $\nu_e$ appearance (left), $\nu_\mu$ disappearance (middle), $\nu_\tau$ appearance (right) considering case-IV, $\Gamma_{31} = \Gamma_{32} = 2.3 \times 10^{-23}~ GeV,~ \Gamma_{21} = 0$}
\label{fig10}
\end{figure*}

We present the oscillation probabilities in fig.~(\ref{fig10}) with $\Gamma_{31}$ and $\Gamma_{32}$ are nonzero and equal. The $P_{\mu e}$ in the left plot is similar to the corresponding plot of case-I and case-III but $P_{\mu \mu}$ (in the middle) and $P_{\mu \tau}$ (in the right) differ. $P_{\mu \mu}$ ($P_{\mu \tau}$) increases (decreases) significantly w.r.t SM for $n \ge 0$ in the energy range $4-9~GeV$. Beyond that both the $P_{\mu \mu}$ and $P_{\mu \tau}$ are lesser than the corresponding SM probability.

\begin{figure*}[hbt!]
\includegraphics[width=0.325\linewidth]{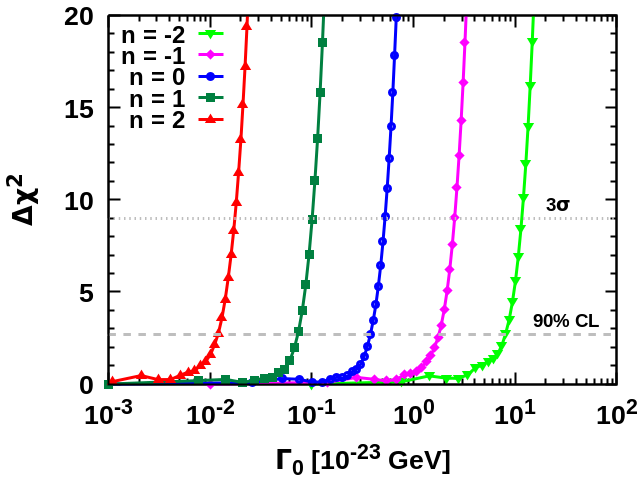}
\includegraphics[width=0.325\linewidth]{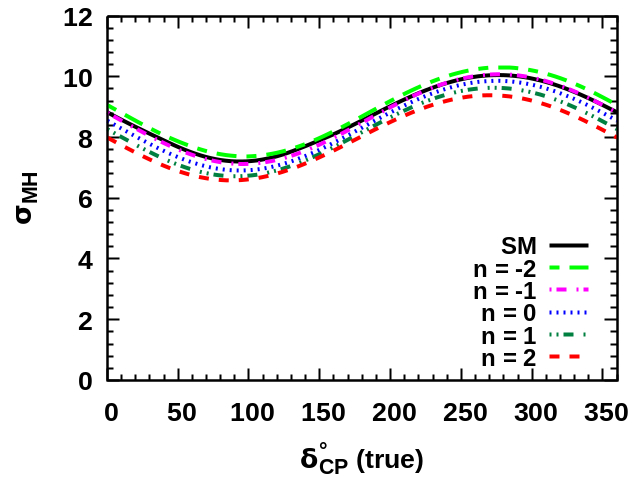}
\includegraphics[width=0.325\linewidth]{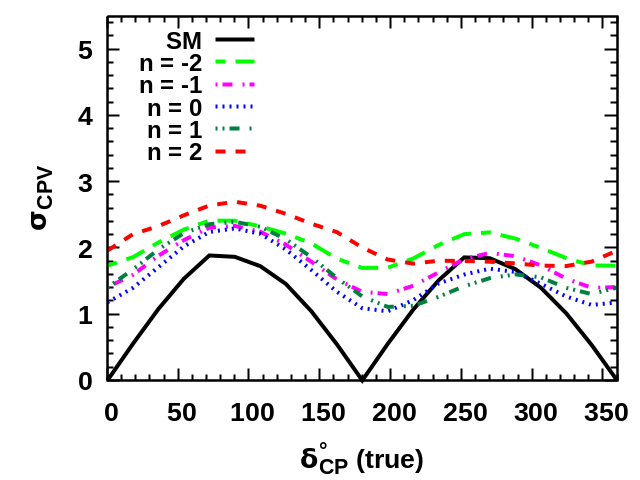}
\caption{Case-IV for P2O experiment (90kW beam). In the left, we present $\Delta \chi^2$ as a function of $\Gamma_0$ (test) based on power law index $n = 0, \pm 1, \pm 2$. Horizontal dashed and dotted lines stands for 90\% and $3\sigma$ CL respectively. In the middle and right panels we show the significance of the experiment to discover MH and CP violation.}
\label{fig11}
\end{figure*}

\begin{figure*}[hbt!]
\includegraphics[width=0.325\linewidth]{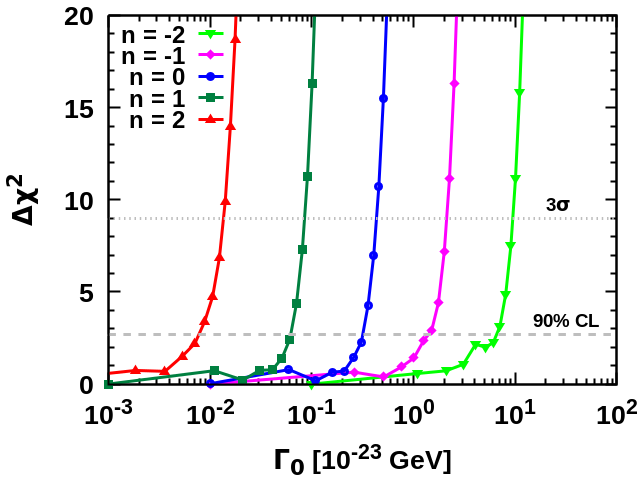}
\includegraphics[width=0.325\linewidth]{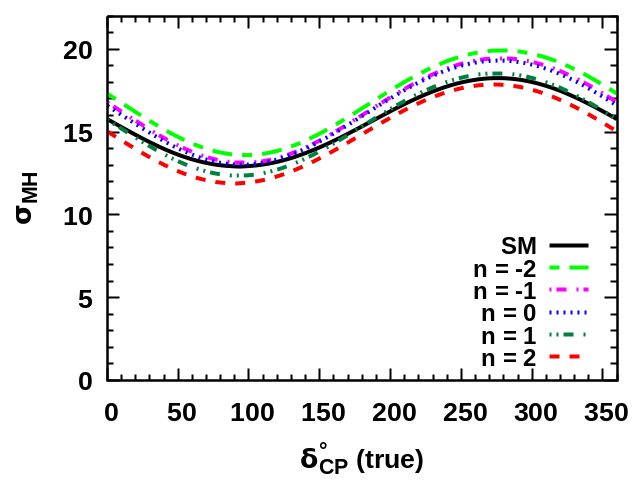}
\includegraphics[width=0.325\linewidth]{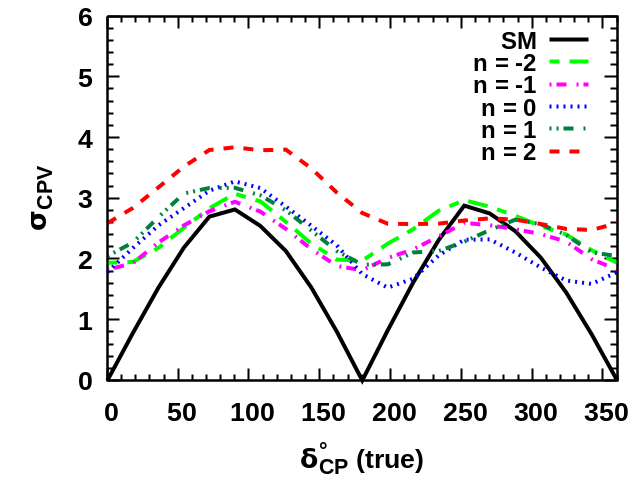}
\caption{Case-IV for P2O-upgrade (450 kW beam power). In the left, we present $\Delta \chi^2$ as a function of $\Gamma_0$ (test) based on power law index $n = 0, \pm 1, \pm 2$. Horizontal dashed and dotted lines stands for 90\% and $3\sigma$ CL respectively. In the middle and right panels we show the significance of the experiment to discover MH and CP violation.}
\label{fig12}
\end{figure*}

Based on the assumption in case-IV, in fig.~(\ref{fig11}) and fig.~(\ref{fig12}) we have shown the bounds on $\Gamma$ in the left panel, the $\sigma_{MH}$ and the $\sigma_{CPV}$ in the middle and the right panels respectively. These plots are similar to the plots in case-I, and hence same inferences can be deduced for this case.
\subsection*{Discussion:}
From fig.~(\ref{fig1}), fig.~(\ref{fig4}), fig.~(\ref{fig7}), fig.~(\ref{fig10}), we observe that varying values of $n$ result in distinct spectral distortions of oscillation probabilities in all the scenarios. It is evident from these figures, that in all the cases the neutrino oscillation probabilities are sensitive to positive powers of $n$. This arises due to the high energy range ($E_\nu \ge E_0$) in eq.~(\ref{G-powerlaw}) of the neutrino beam w.r.t reference value ($E_0 = 1$~GeV) considered. This result aligns with conclusions drawn in ref.~\cite{Romeri:2023cgt} for the accelerator data. From case-I, case-III and case-IV, it can be seen that $P_{\mu e}$ has the same spectral distortion for all the values of $n$. In addition, for $n \geq 0$ there is a peak appearing around 10 GeV for these cases. This can be attributed to the non-zero value of $\Gamma_{32}$ in these cases. The same conclusion that $\Gamma_{32}$ only affects the $P_{\mu e}$ channel can be verified from the left plot of fig.~(\ref{fig4}) belonging to case-II. On the contrary, $\Gamma_{32}$ does not play a significant role in the $P_{\mu\mu}$ channel for the neutrino energy $2 GeV < E_\nu < 12 GeV$. This can be seen from the middle panels of fig.~(\ref{fig4}) and fig.~(\ref{fig7}), where $\Gamma_{32} = 0$ in the former and $\Gamma_{32} \neq 0$ in the latter.

\begin{table}[!htbp]\centering
    \begin{tabular}{|c|c|c|c|c|c|c|c|c|}
\hline\hline \multirow{2}{*}{$\mathrm{n}$} & \multirow{2}{*}{$\mathrm{CL}$} & Case-I & Case-II & Case-III & Case-IV  \\
 & & $\Gamma_{21} = \Gamma_{31} = \Gamma_{32}$ & $\Gamma_{21} = \Gamma_{31}$ & $\Gamma_{21} = \Gamma_{32}$ & $\Gamma_{31} = \Gamma_{32}$ \\
 
\hline \multirow{2}{*}{$n=-2$} & $90 \%$ & $8.06 \times 10^{-23}$ & $6.0 \times 10^{-23}$ & $1.008 \times 10^{-22}$ & $8.0 \times 10^{-23}$ \\
 & $3\sigma$ & $1.15 \times 10^{-22}$ & $1.11 \times 10^{-22}$ & $1.56 \times 10^{-22}$ & $1.15 \times 10^{-22}$ \\

\hline \multirow{2}{*}{$n=-1$} & $90 \%$ & $1.8 \times 10^{-23}$ & $1.36 \times 10^{-23}$ & $2.19 \times 10^{-23}$ & $1.75 \times 10^{-23}$ \\
 & $3\sigma$ & $2.55 \times 10^{-23}$ & $2.46 \times 10^{-23}$ & $3.37 \times 10^{-23}$ & $2.53 \times 10^{-23}$  \\

\hline \multirow{2}{*}{$n = 0$} & $90 \%$ & $3.76 \times 10^{-24}$ & $2.76 \times 10^{-24}$ & $4.26 \times 10^{-24}$ & $3.76 \times 10^{-24}$ \\
 & $3\sigma$ & $5.3 \times 10^{-24}$ & $5.18 \times 10^{-24}$ & $6.7 \times 10^{-24}$ & $5.23 \times 10^{-24}$ \\

\hline \multirow{2}{*}{$n = 1$} & $90 \%$ & $7.4 \times 10^{-25}$ & $5.4 \times 10^{-25}$ & $7.31 \times 10^{-25}$ & $7.3 \times 10^{-25}$  \\
 & $3\sigma$ & $1.04 \times 10^{-24}$ & $1.02 \times 10^{-24}$ & $1.23 \times 10^{-24}$ & $1.008 \times 10^{-24}$ \\

\hline \multirow{2}{*}{$n=2$} & $90 \%$ & $1.24 \times 10^{-25}$ & $9.07 \times 10^{-26}$ & $1.007 \times 10^{-25}$ & $1.2 \times 10^{-25}$  \\
 & $3\sigma$ & $1.79 \times 10^{-25}$ & $1.87 \times 10^{-25}$ & $1.95 \times 10^{-25}$ & $1.74 \times 10^{-25}$  \\

\hline\hline
\end{tabular}
\captionsetup{justification=centering}
 \caption{Bounds obtained in different cases for 90 kW beam power.}
 \label{table:3}
\end{table}
From table-\ref{table:3} and table-\ref{table:4}, it can be observed that case-I and case-IV give rise to similar bounds on $\Gamma$ parameter. Firstly, $\Gamma_{21} \neq 0$ does not effect the $P_{\mu e}$ channel as can be seen from the left panel of fig.~(\ref{fig1}) and fig.~(\ref{fig10}). Additionally, from the middle panel of fig.~(\ref{fig1}) and fig.~(\ref{fig10}) it can be seen that the difference in $P_{\mu \mu}$ values dominant after $E_\nu \ge 8$ GeV. However, the P2O flux gradually falls after $E_\nu \sim 7$ GeV and the analysis window for the $\chi^2$ plots are considered as $2 GeV < E_\nu < 12 GeV$. Consequently, the bounds on $\Gamma$ for these cases (case-I and case-IV) are similar in both P2O and P2O-upgrade experiments. For the case of $n = 0$ (3rd row) in table-\ref{table:3} and table-\ref{table:4} one can note that the bounds on $\Gamma$ from P2O and P2O-upgrade would be better than the bounds reported in ref.~\cite{Romeri:2023cgt}. Specifically, case-II poses strong constraints on $\Gamma_{jk} \leq 2.76 \times 10^{-24}$ GeV for P2O and $\Gamma_{jk} \leq 1.89 \times 10^{-24}$ GeV for P2O-upgrade, which are found to be better than the bounds reported by T2K and MINOS joint analysis in ref.~\cite{gomesalg:2023}, upcoming DUNE experiment in ref.~\cite{Gomes:2019for} and competitive with the recent bounds obtained from neutrinos detected by IceCube in ref.~\cite{Coloma:2018ice, Collab:2024icecube}. For $n = -1, -2$ the bounds on $\Gamma_{jk}$ are $\mathcal{O}(10^{-23})$ at 90\% CL. These bounds are less constrained when compared to the bounds from reactor data of ref.~\cite{Romeri:2023cgt} as expected.

\begin{table}[!t]\centering
    \begin{tabular}{|c|c|c|c|c|c|c|c|c|}
\hline\hline \multirow{2}{*}{$\mathrm{n}$} & \multirow{2}{*}{$\mathrm{CL}$} & Case-I & Case-II & Case-III & Case-IV  \\
 & & $\Gamma_{21} = \Gamma_{31} = \Gamma_{32}$ & $\Gamma_{21} = \Gamma_{31}$ & $\Gamma_{21} = \Gamma_{32}$ & $\Gamma_{31} = \Gamma_{32}$ \\
 
\hline \multirow{2}{*}{$n=-2$} & $90 \%$ & $6.4 \times 10^{-23}$ & $3.9 \times 10^{-23}$ & $6.5 \times 10^{-23}$ & $6.4 \times 10^{-23}$ \\
 & $3\sigma$ & $9.0 \times 10^{-23}$ & $7.6 \times 10^{-23}$ & $1.1 \times 10^{-22}$ & $9.3 \times 10^{-23}$ \\

\hline \multirow{2}{*}{$n=-1$} & $90 \%$ & $1.4 \times 10^{-23}$ & $8.5 \times 10^{-24}$ & $1.5 \times 10^{-23}$ & $1.46 \times 10^{-23}$ \\
 & $3\sigma$ & $2.1 \times 10^{-23}$ & $1.75 \times 10^{-23}$ & $2.5 \times 10^{-23}$ & $2.0 \times 10^{-23}$  \\

\hline \multirow{2}{*}{$n = 0$} & $90 \%$ & $3.1 \times 10^{-24}$ & $1.89 \times 10^{-24}$ & $2.5 \times 10^{-24}$ & $3.1 \times 10^{-24}$ \\
 & $3\sigma$ & $4.2 \times 10^{-24}$ & $4.0 \times 10^{-24}$ & $5.0 \times 10^{-24}$ & $4.7 \times 10^{-24}$ \\

\hline \multirow{2}{*}{$n = 1$} & $90 \%$ & $6.2 \times 10^{-25}$ & $3.6 \times 10^{-25}$ & $4.2 \times 10^{-25}$ & $6.3 \times 10^{-25}$  \\
 & $3\sigma$ & $8.5 \times 10^{-25}$ & $7.7 \times 10^{-25}$ & $8.5 \times 10^{-25}$ & $8.1 \times 10^{-25}$ \\

\hline \multirow{2}{*}{$n=2$} & $90 \%$ & $7.8 \times 10^{-26}$ & $5.87 \times 10^{-26}$ & $6.0 \times 10^{-26}$ & $7.5 \times 10^{-26}$  \\
 & $3\sigma$ & $1.5 \times 10^{-25}$ & $1.4 \times 10^{-25}$ & $1.1 \times 10^{-25}$ & $1.3 \times 10^{-25}$  \\

\hline\hline
\end{tabular}
\captionsetup{justification=centering}
 \caption{Bounds obtained in different cases for 450 kW beam power.}
 \label{table:4}
\end{table}
\section{Conclusions}\label{sec5}
In this work, we have investigated the effect of environmentally induced decoherence on the neutrino oscillations at the upcoming P2O experiment. Considering three flavor oscillation framework, we study the viable phenomenological models of decoherence where we assume all the decoherence parameters are non-zero, (or) two of them are non-zero. In both the scenarios, we consider a power law based relation between the decoherence parameter $\Gamma$ and the neutrino energy $E_\nu$. After a thorough review of the oscillation probabilities we note that $\Gamma_{32}$ plays a significant role on the $\nu_e$ appearance channel. Given that in P2O, the neutrino beam energy $E_\nu > E_0$ (reference $E_0 = 1$ GeV), we observe that when $n \geq 0$ the effect decoherence on the oscillation probabilities is prominent. 

Assuming that there is no decoherence in the true neutrino spectra, we obtain bounds on $\Gamma_{jk}$ parameter for both 90kW (P2O) and 450kW proton beam (P2O-upgrade). For the case of $n = 0$ the bounds obtained on $\Gamma_{jk}$ in both P2O and P2O-upgrade across all the cases would be better than the bounds reported in ref.~\cite{Romeri:2023cgt}. For case-II, we obtain $\Gamma_{jk} \leq 2.76 \times 10^{-24}$ GeV for P2O and $\Gamma_{jk} \leq 1.89 \times 10^{-24}$ GeV for P2O-upgrade, which are found to be the most stringent bounds among the bounds reported from other experiments.

Further, we also explore how the discovery potential of P2O and P2O-upgrade is affected if one assumes non-zero decoherence in the true spectrum and standard three flavour oscillations in the test spectrum.
We note that the comprehensive effect of $P_{\mu e}~,~P_{\mu \mu}$ and the corresponding anti-neutrino probabilities has not significantly affected the MH sensitivity of the experiment. However, when we assume the presence of decoherence in nature, a fake non-zero CP phase causes a rejection in the null hypothesis i.e., $\delta_{CP}=0$, $\pi$ in all the decoherence models. In conclusion, based on our analysis, it is evident that P2O experiment shows significant promise in constraining the environmental decoherence parameters.

{\bf Acknowledgements:} We thank Anatoly Sokolov for fruitful discussions. We acknowledge Dmitry Zaborov for useful information related to P2O flux. One of the authors (K.~N.~Deepthi) would like to thank DST-SERB SIRE program for the financial support.

\bibliographystyle{elsarticle-num}
\bibliography{bibliography.bib}
 
\end{document}